\documentclass[twocolumn,aps,showpacs,showkeys,prb,tightenlines,amsmath,amssymb]{revtex4}
\usepackage{graphicx}
\usepackage{amssymb}
\usepackage{dcolumn}
\usepackage{amsmath}
\usepackage{pifont}
\usepackage{bm}
\usepackage{colordvi}
\newcommand{\bgreek}[1]{\mbox{\boldmath$#1$\unboldmath}}

\begin{document}

\title{Intrinsic electron spin relaxation due to the D'yakonov-Perel' 
mechanism in monolayer MoS$_2$}
\author{L. Wang}
\affiliation{Hefei National Laboratory for Physical Sciences at
Microscale and Department of Physics,
University of Science and Technology of China, Hefei,
Anhui, 230026, China}
\author{M. W. Wu\footnote{Electric address: mwwu@ustc.edu.cn}}
\affiliation{Hefei National Laboratory for Physical Sciences at
Microscale and Department of Physics, University of Science and
Technology of China, Hefei, Anhui, 230026, China}

\date{\today}

\begin{abstract}
Intrinsic electron spin relaxation due to the D'yakonov-Perel' mechanism is studied in monolayer Molybdenum Disulphide.
An intervalley in-plane spin relaxation channel
is revealed due to the opposite effective magnetic fields perpendicular 
to the monolayer Molybdenum Disulphide 
plane in the two valleys together with the intervalley electron-phonon scattering. The intervalley electron-phonon 
scattering is always in the weak scattering limit, 
which leads to a rapid decrease of the in-plane 
spin relaxation time with increasing temperature. A decrease of the in-plane spin
relaxation time with the increase of the electron density is also
shown.  
\end{abstract}

\keywords{Spin relaxation \ Monolayer MoS$_2$}
\pacs{72.25.Rb, 81.05.Hd, 71.10.-w, 71.70.Ej}


\maketitle
\section{INTRODUCTION}
Transition-metal dichalcogenides, similar to graphite, are layered materials 
with weak interlayer van der Waals interaction. These materials can be exfoliated to 
single to few-layer samples as the fabrication of graphene.\cite{novoselov102,ayari101,matte122,radisavljevic6}  
Among these samples, monolayer Molybdenum Disulphide (MoS$_2$) has received much attention 
due to its distinctive properties.\cite{radisavljevic6,splendiani10,lee,mak105,eda11,zhu84,korn99,ellis99,behnia,zeng7,mak7,
wang7,cao3,kim3,xiao108,yun85,kaasbjerg,chei,sallen86,kong101,lembke6,kadantsev152,liu,buscema13,lu110,kosmider87,
kaasbjerg2,ochoa,hshi,xli,song,rostami,zahid,kormanyos,klinovaja,cappelluti} 
It has a direct gap at the inequivalent K and K$^{\prime}$ points,\cite{splendiani10,kaasbjerg,xli,zhu84,chei,hshi} 
which makes it attractive as a two-dimensional channel material in field-effect transistors.
Very recently, the field-effect transistor devices based on monolayer MoS$_2$ have been realized in the experiments 
with high on-off ratio.\cite{radisavljevic6,liu} In addition, 
space inversion symmetry is broken in monolayer  
MoS$_2$ since it is a two-dimensional hexagonal lattice 
consisted of two different sublattices, i.e., Mo and S atoms. 
The absence of the space inversion symmetry results in a valley-dependent 
optical selection rule for interband 
transitions, which allows the realization 
of the valley polarization by optical pumping with circularly polarized 
light.\cite{cao3,mak7,zeng7,behnia,xiao108,zhu84,sallen86} 
Space inversion symmetry breaking can also induce  spin splitting of both 
the conduction and valence bands where the one of the 
valence band is much larger than that of the conduction 
band.\cite{zhu84,ochoa,xiao108,kormanyos,zahid,chei,kadantsev152,kosmider87} 
This spin splitting is essential for spin physics and 
spintronic applications. All these intriguing features make 
monolayer MoS$_2$ of particular interest.

As spin relaxation is crucial for possible realistic spintronic applications, a 
thorough understanding of the spin relaxation time (SRT) 
in monolayer MoS$_2$ is essential. Very recently, Ochoa and
 Rold\'{a}n\cite{ochoa} investigated the intravalley spin-orbit 
mediated spin relaxation in 
monolayer MoS$_2$. For electrons, they 
calculated the out-of-plane spin relaxation due to the extrinsic Rashba spin-orbit 
coupling (SOC) induced by an external out-of-plane electric field and also 
the in-plane one due to the intrinsic SOC from the 
contribution of the valence band due to the space inversion symmetry. 
However, according to the latest report by
 Korm\'{a}nyos {\it et al.},\cite{kormanyos} 
this intrinsic SOC from the contribution of the valence band is weak 
since the splitting of the valence band is much smaller 
than the typical band gap. 
In contrast, the intrinsic SOC from the conduction band, 
which manifests itself as a Zeeman-like term with opposite effective magnetic fields 
in the two valleys, is absent in their calculation.\cite{ochoa} The Zeeman-like term 
can give rise to an intervalley spin relaxation
channel in the presence of intervalley electron-phonon scattering, which has been shown of significant importance to the 
spin relaxation in both rippled single-layer graphene\cite{pzhang112} and 
also bilayer graphene.\cite{lwang} 
Moreover, only the disorder is included in a phenomenological manner 
in their calculation. The electron-electron Coulomb and electron-phonon 
interaction, which have been demonstrated to play an important role in spin relaxation in 
semiconductors\cite{wureview} and also graphene at high temperature,\cite{yzhou}  
are absent.

In the present work, with the electron-electron Coulomb, 
(both the intra- and inter-valley) electron-phonon and the
 electron-impurity scatterings 
explicitly included, we study the electron spin relaxation due to 
the D'yakonov-Perel'\cite{dp}  mechanism 
in monolayer MoS$_2$ in the absence of external electric field by 
the kinetic spin Block equation (KSBE) approach.\cite{wureview} 
According to the latest report by 
Korm\'{a}nyos {\it et al.},\cite{kormanyos} 
the effective magnetic field of the intrinsic SOC of the conduction
 band is given by 
\begin{equation}
{\bf \Omega}^{\mu}=2\omega\mu\hat{\bf z}
\label{soc}
\end{equation}
with the $z$-axis being perpendicular to the monolayer MoS$_2$ plane. 
Here, $\omega$ and $\mu=1(-1)$ represent 
the strength of the SOC and K(K$^{\prime}$) valleys, respectively.\cite{kormanyos} 
It is noted that the contribution of the hybridization by 
the intrinsic SOC of the valence band to 
the energy bands is neglected since the energy scale of this spin splitting is much smaller than 
the typical band separation.\cite{kormanyos}
This effective magnetic field is momentum independent, indicating an absence of 
the intravalley inhomogeneous broadening\cite{wuning} for in-plane spins. Therefore, the intravalley spin 
relaxation process contributed by the electron-electron Coulomb, intravalley electron-phonon and electron-impurity 
scatterings are irrelevant to the in-plane spin relaxation. Here, only the intervalley process
contributes to the in-plane spin relaxation. The underlying physics is the same
as the case of bilayer graphene\cite{lwang} and also rippled single-layer graphene.\cite{pzhang112} 
This intervalley spin relaxation channel suppresses the in-plane SRT effectively at high temperature. 
We also find that this system is always in the weak intervalley scattering limit, 
which determines that the in-plane spin relaxation time is proportional to 
the intervalley momentum scattering time. As a result, a monotonic decrease of the in-plane SRT 
with increasing temperature is shown. We also find that
the in-plane SRT decreases with the increase of the electron
density. Moreover, we show a decrease of the in-plane SRT 
with increasing initial spin polarization at low temperature, which is very different from the previous 
studies in both semiconductors\cite{weng,stich} and single-layer graphene\cite{yzhou} but similar to 
the case of bilayer graphene.\cite{lwang}

\section{MODEL AND KSBEs}
With the effective mass approximation, the Hamiltonian of the conduction band 
near the K(K$^{\prime}$) points in monolayer MoS$_2$ can be described by
\begin{eqnarray}
H_{\rm eff}^{\mu}&=&\epsilon_{\mu{\bf k}}+{\bf \Omega}^{\mu}\cdot{\bgreek \sigma}/2,\label{hamil}
\end{eqnarray}
according to the latest report by Korm\'{a}nyos {\it et al.}.\cite{kormanyos}
Here, $\epsilon_{\mu{\bf k}}=\hbar^2{\bf k}^2/(2m^*)$ with ${\bf k}$ and $m^*$ being the in-plane momentum 
relative to the K(K$^{\prime}$) points and the effective mass, respectively. ${\bgreek \sigma}$ are the Pauli matrices 
and ${\bf \Omega}^{\mu}$ is given in Eq.~(\ref{soc}).

We then construct the microscopic KSBEs\cite{wureview} to investigate the 
intrinsic electron spin relaxation in monolayer 
MoS$_2$. The KSBEs are given by\cite{wureview} 
\begin{eqnarray}
\partial_t\hat{\rho}_{\mu{\bf k}}=\partial_t\hat{\rho}_{\mu{\bf k}}|_{\rm coh}+\partial_t\hat{\rho}_{\mu{\bf k}}|_{\rm scat},
\label{KSBE}
\end{eqnarray}  
in which $\hat{\rho}_{\mu{\bf k}}$ stand for the density matrices of electrons
with the diagonal terms $\rho_{\mu{\bf k},\sigma\sigma}\equiv f_{\mu{\bf k}\sigma}\ (\sigma=\pm \frac{1}{2})$
representing the distribution functions and the off-diagonal ones $\rho_{\mu{\bf k},(\frac{1}{2})(-\frac{1}{2})}
=\rho_{\mu{\bf k},(-\frac{1}{2})(\frac{1}{2})}^*$ describing the spin
coherence. The coherent terms $\partial_t\hat{\rho}_{\mu{\bf k}}|_{\rm
    coh}$ are given in Ref.~\onlinecite{yzhou}. 
$\partial_t\hat{\rho}_{\mu{\bf k}}|_{\rm scat}$ are the scattering terms 
including the electron-electron Coulomb
($|V_{{\bf k},{\bf k}^{\prime}}^{\mu}|^2$), 
electron-impurity ($|U_{{\bf k},{\bf k}^{\prime}}^{\mu}|^2$), intravalley
electron-acoustic  phonon, electron-optical phonon,
 and especially the intervalley electron-phonon scattering
including the electron-KTA phonon, electron-KLA phonon,
electron-KTO phonon and electron-KLO phonon  scatterings.
Here, KTA, KLA, KTO and KLO correspond to the transverse acoustic, longitudinal 
acoustic, transverse optical and longitudinal optical phonon modes at K point, 
respectively.\cite{kaasbjerg,xli} 
The detailed expressions of the above scattering terms can be found in
  Ref.~\onlinecite{yzhou} (Note that the form factor in the electron-impurity
  and electron-electron Coulomb scatterings in Ref.~\onlinecite{yzhou} is absent
  here). The scattering matrix elements 
\begin{eqnarray}
|V_{{\bf k},{\bf {k-q}}}^{\mu}|^2=\Big(\frac{V_{\bf q}^{(0)}}{\varepsilon({\bf q})}\Big)^2\label{eematrix}
\end{eqnarray}
and 
\begin{eqnarray}
|U_{{\bf k},{\bf k-q}}^{\mu}|^2=Z_i^2|V_{{\bf k},{\bf {k-q}}}^{\mu}|^2\label{eimatrix}
\end{eqnarray}
with $Z_i$ being the impurity charge number (assumed to be 1 in our calculation). 
The quantity
\begin{eqnarray}
\varepsilon({\bf q})=1-V_{\bf q}^{(0)}\sum_{\mu{\bf k}s}\frac{f_{{\bf k}s}^{\mu}-f_{{\bf k+q}s}^{\mu}}
{\epsilon_{\mu{\bf k}}-\epsilon_{\mu{\bf k+q}}}\label{screening}
\end{eqnarray}
stands for the screening under the random phase approximation.\cite{mahan,haug} 
$V_{\bf q}^{(0)}=2\pi e^2/(\kappa q)$ is
the two-dimensional bare Coulomb potential with $\kappa$ being the relative static dielectric 
constant.\cite{chei} The electron-phonon scattering matrix elements 
are explicitly given by the latest reports by Kaasbjerg {\it et al.}.\cite{kaasbjerg,kaasbjerg2} 
Specifically, we lay out the matrix elements of the electron-KTA ($|M^{\rm
    KTA}_{{\mu{\bf k}},{\mu^{\prime}{\bf k}^{\prime}}}|^2$) 
and -KLO ($|M^{\rm KLO}_{{\mu{\bf k}},{\mu^{\prime}{\bf k}^{\prime}}}|^2$) 
phonon scatterings, which 
play a dominant role in the in-plane spin relaxation as will be shown later. 
\begin{eqnarray}
|M^{\rm KTA}_{{\mu{\bf k}},{\mu^{\prime}{\bf k}^{\prime}}}|^2&=&\frac{\hbar^2(D^1_{\rm K,TA})^2q^2}
{2A\rho\Omega_{\rm K,TA}}\delta_{\mu^{\prime},-\mu},\label{KTA}\\
|M^{\rm KLO}_{{\mu{\bf k}},{\mu^{\prime}{\bf k}^{\prime}}}|^2&=&\frac{\hbar^2(D^0_{\rm K,LO})^2}
{2A\rho\Omega_{\rm K,LO}}\delta_{\mu^{\prime},-\mu},\label{KLO}
\end{eqnarray}
where $A$ is the area of the sample; $\rho$ is the mass density of the monolayer MoS$_2$; 
$\Omega_{\rm K,TA}$ and $\Omega_{\rm K,LO}$ are the energies of KTA and KLO phonon modes, respectively; 
$D^1_{\rm K,TA}$ ($D^0_{\rm K,LO}$) is the first- (zeroth-) order 
deformation potential corresponding to the electron-KTA (-KLO) phonon scattering; 
$q=|{\bf k}-{\bf k}^{\prime}|$. It is noted that the screening by the carriers is included in the electron-phonon scattering in their works with 
the random phase approximation for acoustic phonon and
  Debye-H\"{u}ckel one in the nondegenerate limit for other phonon 
modes.\cite{mahan,haug} It is
further noted that they pointed out that the screening can be neglected for the
intervalley electron-phonon scattering due to the large wave vectors of
phonons.\cite{kaasbjerg} Therefore, in the present work, we only include the screening for the
intravalley electron-phonon scattering with the random phase approximation.\cite{mahan,haug}

\section{NUMERICAL RESULTS}
In the literature, there still remain controversies over the band structure of monolayer MoS$_2$,\cite{xli,kaasbjerg,
kormanyos,zahid,chei,splendiani10,kadantsev152,hshi} where different energy gaps, effective masses and the 
spin splittings are given. Here, 
we take the effective mass and the strength of the SOC being $m^*=0.48m_0$\cite{kaasbjerg,kormanyos} and 
$\omega=1.5$\ meV,\cite{kormanyos,kadantsev152,kosmider87} respectively. $m_0$ stands for the free electron mass. The relative static dielectric constant is 
chosen to be $\kappa=3.43$.\cite{chei} The mass density $\rho=3.1\times 10^{-7}\ $g/cm$^2$; the 
KTA and KLO phonon energies are $\Omega_{\rm K,TA}=23\ $meV and $\Omega_{\rm K,LO}=42\ $meV, respectively; 
the deformation potentials for the KTA and KLO phonons are $D^1_{\rm K,TA}=5.9\ $eV 
and $D^0_{\rm K,LO}=2.6\times 10^8\ $eV/cm, separately.\cite{kaasbjerg}
With these parameters, our results are obtained by numerically solving the KSBEs [Eq.~(\ref{KSBE})].\cite{wureview,kikkawa} 
The initial spin polarization is set to be $2.5$~\% and the spin-polarization 
direction is chosen in the monolayer MoS$_2$ plane unless otherwise specified. 
As reported, in contrast to the out-of-plane spin 
orientation,\cite{cao3,mak7,zeng7,behnia,xiao108,zhu84,sallen86} 
the in-plane one is difficult to generate optically due to 
the large out-of-plane spin splitting of the valence band. However, with an
in-plane magnetic field, the in-plane component 
of the spin polarization can be obtained from the 
out-of-plane spin orientation.\cite{stich76} In
addition, the in-plane spins can also be realized electrically by 
ferromagnetic contacts.\cite{stephens,tombros} 
The electrical method has been
widely used in spin transport measurements in
 semiconductors\cite{stephens} and also graphene.\cite{tombros}

\subsection{Temperature dependence of spin relaxation}
\begin{figure}[bth]
\centering
\includegraphics[width=5.4cm]{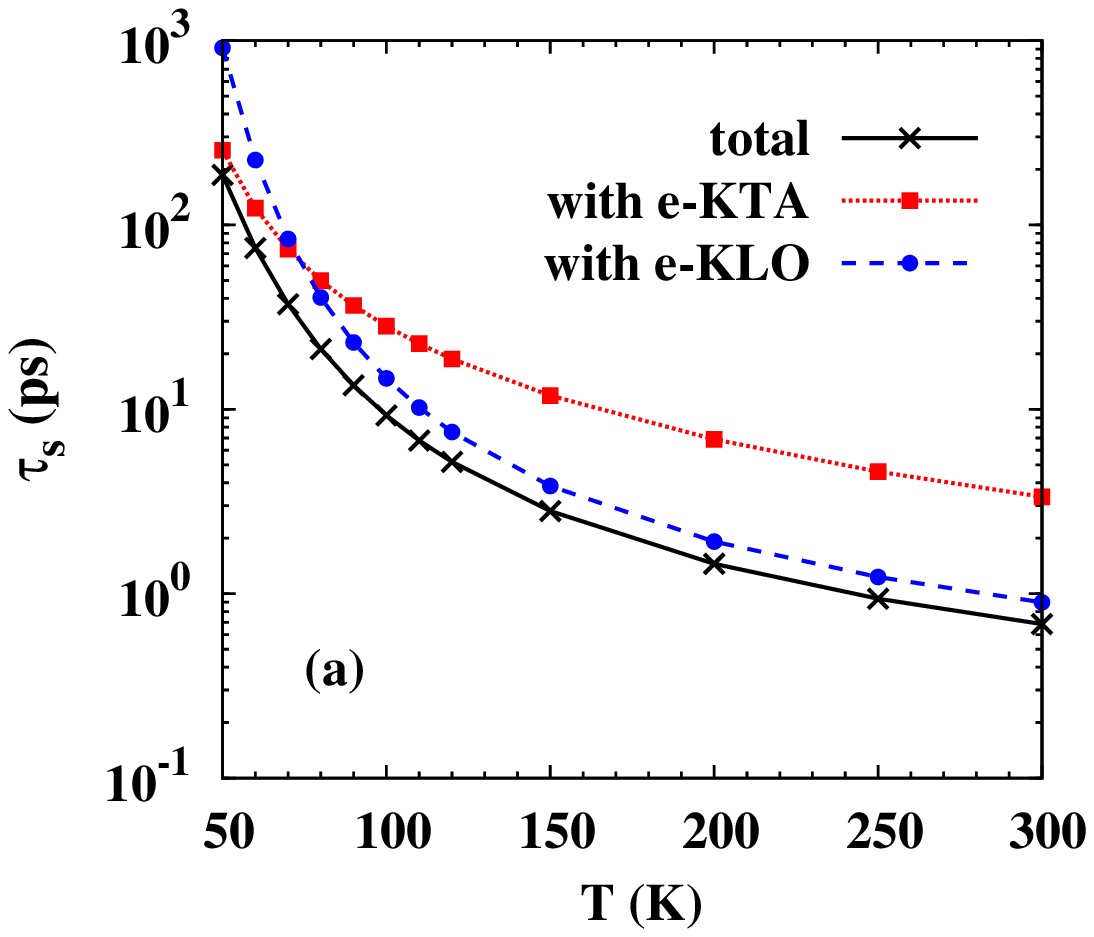}
\includegraphics[width=5.4cm]{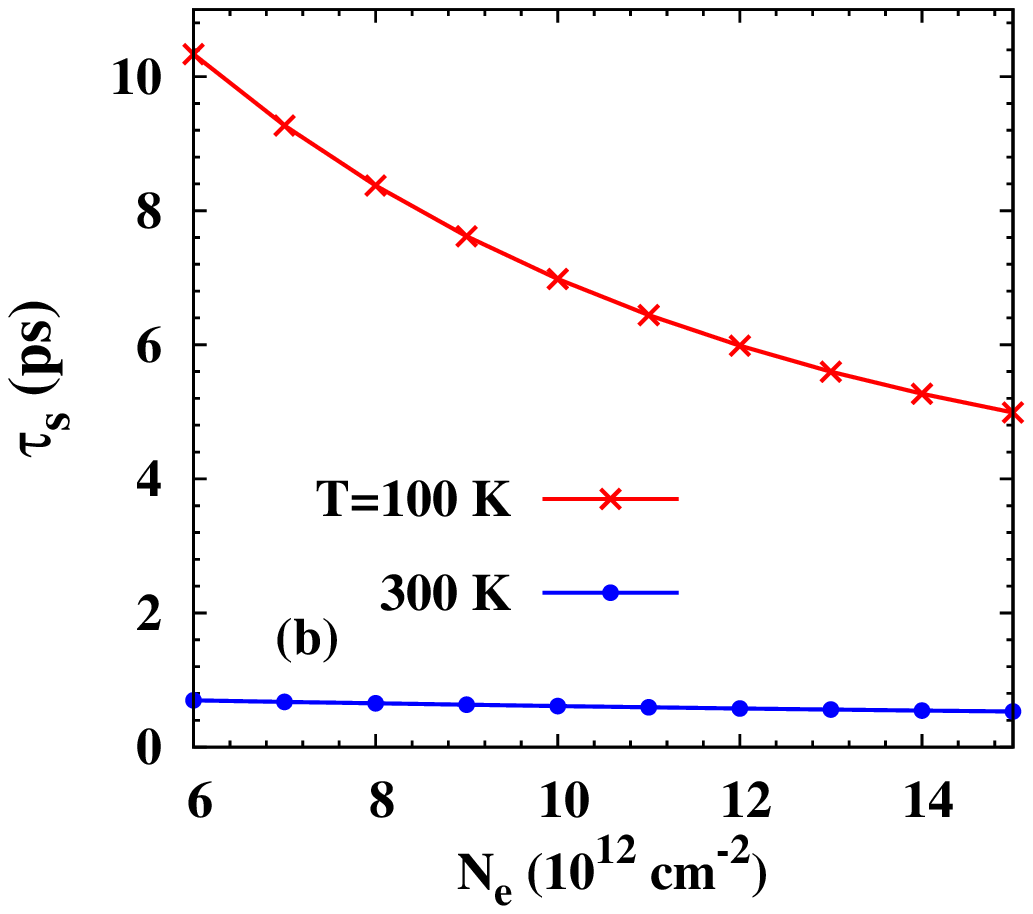}
\includegraphics[width=5.4cm]{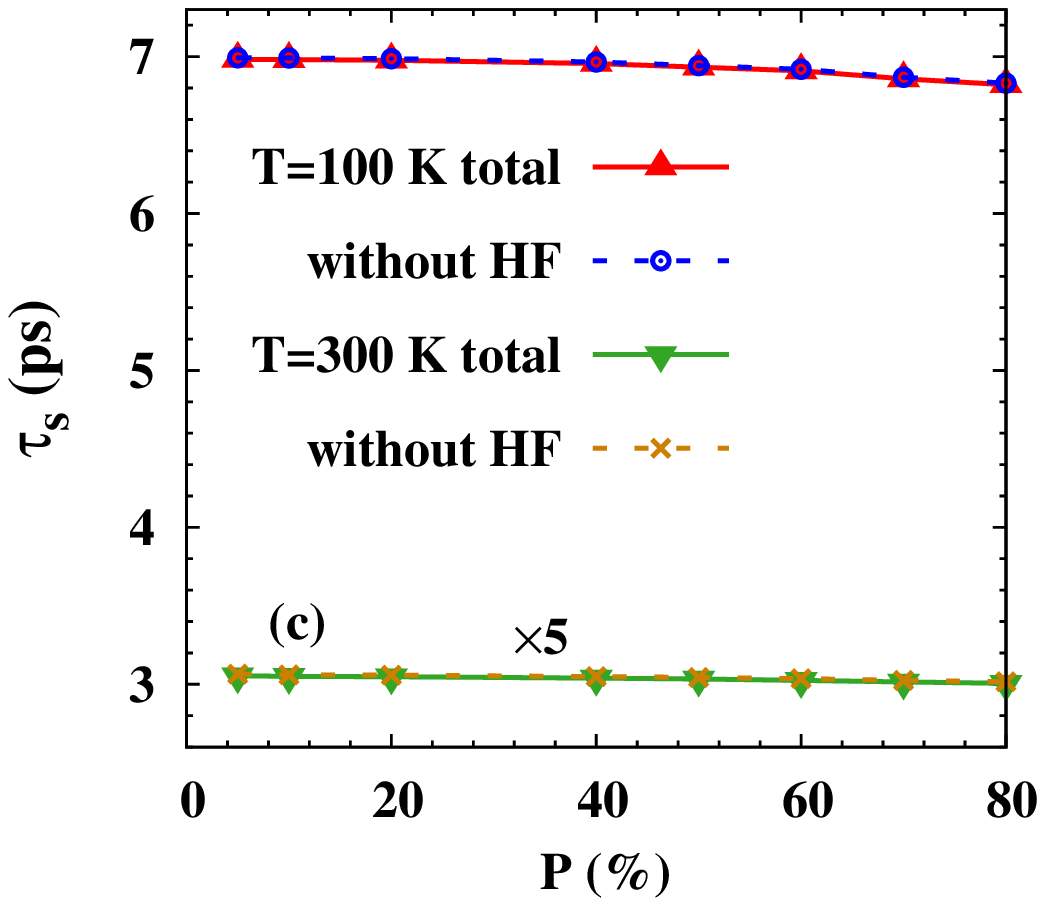}
\caption{(Color online) (a) Total in-plane SRT $\tau_{s}$ ($\times$)
and that calculated with only the electron-KTA phonon 
($\blacksquare$) or electron-KLO phonon ($\bullet$) 
scattering included as function of temperature $T$ with the electron density 
$N_e=7\times 10^{12}\ $cm$^{-2}$; (b) $\tau_{s}$ as a function of $N_e$ at $T=100\ (300)\ $K; (c)
Total in-plane SRT $\tau_{s}$ $\blacktriangle$ ($\blacktriangledown$)
and that calculated without the Coulomb HF term $\circ$ ($\times$) as function of the initial spin polarization 
$P$ at $T=100\ (300)\ $K with $N_e=10^{13}\ $cm$^{-2}$.}
\label{fig1}
\end{figure}

We first study the temperature dependence of the spin relaxation.
The in-plane SRT $\tau_{s}$ as function of temperature $T$ 
is plotted in Fig.~\ref{fig1}(a). It is seen that the SRT decreases monotonically with the increase of 
the temperature. To understand this behavior, we calculate the temperature dependence of the SRT 
with only the electron-electron Coulomb, electron-impurity, intravalley 
electron-acoustic phonon, electron-optical phonon,
 or intervalley electron-phonon scattering included, separately. 
We find that the 
contributions of the electron-electron Coulomb, electron-impurity, 
intravalley electron-phonon scatterings, which only 
influence the intravalley spin relaxation channel, are negligible
 due to the absence of a momentum-dependent 
effective magnetic field of the SOC [Eq.~(\ref{soc})].\cite{wureview} 
Therefore, the SRT is only determined by the intervalley spin relaxation 
channel.\cite{pzhang112,lwang} With the 
Zeeman-like term being $2\omega\mu\hat{\bf z}$ [Eq.~(\ref{soc})], 
the in-plane SRT is given by
\begin{eqnarray}
\tau_{s}=\left\{
\begin{array}{ll}
\tau_v ~\mbox{{\rm \ \ weak\ scattering}}\ (2\omega\tau_v\ge 1)\\
\frac{1}{2\omega^2\tau_v} ~\mbox{{\rm \ \ \ strong scattering}}\ 
(2\omega\tau_v\ll 1),
\end{array}\right.\label{formulazhang}
\end{eqnarray}
according to the report by Zhang {\it et al.}\cite{pzhang112} based on the 
elastic scattering approximation. Here, $\tau_v$ represents 
the intervalley electron-phonon scattering time. 
It is noted that the intervalley electron-phonon scattering is always in the weak 
scattering limit for the electron density upto $2\times 10^{13}\ $cm$^{-2}$, 
therefore $\tau_{s}=\tau_v$ throughout our work. 
As a result, the SRT decreases with the enhancement of the intervalley electron-phonon 
scattering as the temperature increases. 

In addition, we also show a comparison of contributions from 
two leading intervalley electron-phonon scatterings, i.e., electron-KTA and electron-KLO ones in Fig.~\ref{fig1}(a). 
We find that the electron-KTA phonon scattering plays a more important role in spin relaxation 
at low temperature whereas the electron-KLO phonon scattering becomes more important at high temperature. 
This can be understood from the contribution of the electron-phonon scattering matrix element 
together with phonon number. It is noted that the electron-KTA phonon scattering matrix element is smaller 
than the electron-KLO phonon one whereas the KTA phonon mode has a larger 
phonon number due to a smaller phonon energy.\cite{kaasbjerg,xli} 
At low temperature, the phonon number of KTA mode is much larger than that of KLO mode, which 
makes the electron-KTA phonon scattering stronger and hence more important to the spin relaxation. However, 
at high temperature, the phonon numbers of two modes become comparable. Then the electron-KLO phonon 
scattering plays a more important role in spin relaxation due to the larger scattering matrix element.  
It is noted that the contributions of the electron-KLA and -KTO phonon scatterings to the 
spin relaxation are marginal.

\subsection{Electron-density dependence of spin relaxation}
Then we turn to investigate the electron-density dependence of the in-plane SRT 
and show the results at $T=100\ (300)\ $K in
Fig.~\ref{fig1}(b). We find that the SRT decreases rapidly with increasing 
electron density at $T=100\ $K but decreases mildly at $T=300\
  $K. We first focus on the case at $T=100\ $K. The SRT is dominated by the
  electron-KTA and -KLO phonon scatterings as pointed out in the temperature
  dependence of the spin relaxation. The electron-KLO phonon scattering, which
  has a momentum-independent matrix element [see Eq.~(\ref{KLO})], is
  insensitive to the electron density. However, the electron-KTA phonon
  scattering is enhanced with increasing electron density as its matrix element 
[see Eq.~(\ref{KTA})] increases with the increase of the momentum. Therefore,
when the electron density increases, the enhancement of the electron-KTA phonon scattering leads to the decrease of the
  SRT ($\tau_{s}=\tau_v$). As for the case at $T=300\ $K, the increase of the matrix element of the electron-KTA
  phonon scattering becomes insensitive. In addition, the
  electron-KLO phonon scattering plays a more important role in spin relaxation than that in the case at
  $T=100\ $K. Both lead to a mild decrease of the SRT.

\subsection{Initial spin polarization dependence of spin relaxation}
In the previous investigations in both 
semiconductor systems\cite{stich,weng} and
single-layer graphene,\cite{yzhou} a significant increase of the SRT is 
shown with the increase of the initial spin polarization, which originates from 
the contribution of the Coulomb Hartree-Fock (HF) term. However, in 
the case of bilayer graphene, the SRT decreases rapidly with increasing initial spin 
polarization at low temperature whereas shows a mild increase at high temperature.\cite{lwang} 
In this case, the SRT is determined by the intervalley electron-phonon scattering.
Here, we also investigate the initial spin polarization dependence of spin relaxation. 
In Fig.~\ref{fig1}(c), we plot the in-plane SRT as function of the initial spin 
polarization $P$ at $T=100\ (300)\ $K. 
We find that the SRT shows a decrease with increasing $P$ at $T=100\ $K 
whereas becomes insensitive to $P$ at $T=300\ $K. We first focus on 
the case of $T=100\ $K. The decrease of the SRT at $T=100\ $K is
very different from the previous studies in both 
semiconductors\cite{stich,weng} and
single-layer graphene\cite{yzhou} but similar to the case of bilayer graphene 
at low temperature.\cite{lwang} 
The underlying physics is understood as follows. 
The SRT is also determined by the intervalley electron-phonon
scattering whereas the contribution of the Coulomb HF term is negligible to the spin relaxation 
by comparing the calculation with and without the Coulomb HF term shown in Fig.~\ref{fig1}(c). 
The intervalley electron-phonon scattering is in the weak intervalley scattering limit, i.e., the SRT 
$\tau_{s}=\tau_v$ [Eq.~(\ref{formulazhang})]. As a result, 
the SRT decreases with the increase of the initial spin polarization 
due to the enhancement of the intervalley electron-phonon scattering.\cite{lwang} 
As for the case of $T=300\ $K, the SRT is also
determined by $\tau_{s}=\tau_v$ [Eq.~(\ref{formulazhang})]. The
  insensitive initial spin polarization dependence of $\tau_s$ originates from
  that of $\tau_v$ at $T=300\ $K.\cite{lwang}

\section{DISCUSSION and Summary}
As mentioned previously, the effective magnetic field [see Eq.~(\ref{soc})] is momentum independent, 
which leads to the absence of the intravalley relaxation 
channel for in-plane spins. 
This effective magnetic field can become  momentum dependent 
when the contribution of the hybridization by the intrinsic SOC 
of the valence band to the energy bands is taken into account.\cite{kormanyos} 
This intravalley inhomogeneous broadening together with the
intravalley scattering opens an intravalley spin relaxation
channel. The intravalley process competes 
with the intervalley one discussed here and the competition between these two
processes needs further investigations.\cite{ochoa} In addition, in real systems, 
the effective magnetic field can be sensitive to the random local
strains\cite{luttrell,sun79} due to the symmetry breaking, which also results in an intravalley inhomogeneous
broadening and hence the intravalley spin relaxation process.
However, this is out of the scope of the present work 
since the effect of the random local strains on 
the spin-orbit coupling is still unclear in monolayer MoS$_2$.

In summary, we have investigated the intrinsic 
electron spin relaxation due to the D'yakonov-Perel' mechanism
in monolayer MoS$_2$. The
effective magnetic field of the intrinsic SOC is momentum independent, which
supplies a Zeeman-like term with opposite effective magnetic fields
perpendicular to the mononlayer MoS$_2$ plane in the two valleys. Since it is
independent on momentum, the intravalley spin relaxation channel is absent for
the in-plane spins. However, the Zeeman-like term, together with the intervalley
electron-phonon scattering, gives rise to an intervalley relaxation channel for in-plane
spin polarization, which is similar to the case in rippled single-layer graphene
and also bilayer graphene. This effect 
has not been yet reported in the literature on
monolayer MoS$_2$.  The intervalley spin relaxation channel 
can markedly suppress the in-plane SRT at high temperature. 
In addition to the intervalley 
electron-phonon scattering, the electron-impurity, 
electron-electron Coulomb and intravalley electron-phonon scatterings are also included to calculate the in-plane SRT. 
However, their contribution to in-plane spin relaxation is negligible 
due to the absence of intravalley relaxation process. 
Moreover, we find that the intervalley scattering is always in the 
weak scattering 
limit in this material. Therefore, the SRT is always proportional to the 
intervalley momentum scattering time. This results in a monotonic 
decrease of the 
in-plane SRT with the increase of temperature. A decrease of the in-plane SRT is also
observed in the electron density dependence. In addition, we find that
the in-plane SRT decreases with the increase of the 
initial spin polarization at low temperature, which is very different from the previous studies 
in both semiconductors and single-layer graphene but similar to the case of bilayer graphene.

\begin{acknowledgments}
This work was supported by the National Natural Science Foundation of
China under Grant No.\ 11334014, the National Basic Research Program of
China under Grant No.\ 2012CB922002 and the Strategic 
Priority Research Program of the
Chinese Academy of Sciences under Grant No.\ XDB01000000. 
\end{acknowledgments}

\end{document}